\documentclass[12pt,draftcls,onecolumn]{IEEEtran}
\usepackage{graphics}
\usepackage{graphicx}
\usepackage{epsfig}

\usepackage{amssymb}
\usepackage{amsthm}

\usepackage{enumerate}
\usepackage{color}
\usepackage{cases}

\def\rm{\mathrm}
\newtheorem{assumption}{Assumption}
\newtheorem{lemma}{Lemma}
\newtheorem{definition}{Definition}
\newtheorem{theorem}{Theorem}
\newtheorem{remark}{Remark}

\begin{document}

\title{Designing Distributed Fixed-Time Consensus Protocols for Linear Multi-Agent Systems Over Directed Graphs}

\author{
Yu~Zhao,  Yongfang Liu,
        %Guanghui Wen,  \emph{Member, IEEE}, and
         Guanrong Chen
        % <-this % stops a space
%\thanks{This work is supported by the National Science Foundation of China under Grant 61225013.}
\thanks{Y. Zhao and Y. Liu are with the School of Automation, Northwestern Polytechnical University, Xi'an Shaanxi, 710129, China
(e-mail: yuzhao5977@gmail.com; liuyongfangpku@gmail.com).}% <-this % stops a space

%\thanks{G. Wen is with the Department of Mathematics, Southeast University,
% Nanjing 210096, China (e-mail: wenguanghui@gmail.com). }% <-this % stops a space
%
%
%
\thanks{G. Chen is with the Department of Electronic Engineering, City University
of Hong Kong, Kowloon, Hong Kong (e-mail: gchen@ee.cityu.edu.hk). }% <-this % stops a space
%\thanks{Copyright (c) 2014 IEEE. Personal use of this material is permitted.
%However, permission to use this material for any other purposes must be
%obtained from the IEEE by sending an email to pubs-permissions@ieee.org}
}

%\title{Consensus of Multi-Agent Systems
%with General Linear and Lipschitz Nonlinear Dynamics Using Distributed Adaptive Protocols}
%\author{ Zhongkui~Li, Wei~Ren,~\IEEEmembership{Member,~IEEE}, Xiangdong~Liu, and Mengyin~Fu%
%\thanks{Z. Li, X. Liu and M. Fu are with the School of
%Automation, Beijing Institute of Technology, Beijing 100081, China (E-mail: zhongkli@gmail.com).}
%\thanks{W. Ren is with the Department of
%Electrical and Computer Engineering, Utah State University, UT
%84322, USA (E-mail: wei.ren@usu.edu).}}

\maketitle
\begin{abstract}
This technical note addresses the distributed fixed-time consensus
protocol design problem for multi-agent systems with general
linear dynamics over directed communication graphs. By using motion planning approaches, a class of distributed fixed-time consensus algorithms are developed, which rely only on the sampling information at some sampling instants. For linear multi-agent systems,  the proposed algorithms solve the fixed-time consensus problem for any directed graph containing a directed spanning tree. In particular, the settling time can be off-line pre-assigned according to task requirements. Compared with the existing results for multi-agent systems, to our best knowledge, it is the first-time to solve fixed-time consensus problems for general linear multi-agent systems over directed graphs having a directed spanning tree.
Extensions to the fixed-time formation flying are further studied for multiple satellites described by Hill equations.
\end{abstract}


% \fi
\IEEEpeerreviewmaketitle

\section{Introduction}
Over the past few years, the coordinative control problems of multi-agent systems are of great interest
to various scientific and engineering communities \cite{Olfati04,Ren05,Cao08,Hong08,Li10}, due to its broad applications in such fields as
spacecraft formation flying, distributed sensor network, automated highway systems,
and so forth \cite{Li13,Liu152,Chenfei15,Zhaornc16}. Compared to the traditional monolithic systems, the coordination control
reduces the systems cost, breaches the size constraints and prolongs the life span of the systems. One interesting and important issue arising from coordination control of multi-agent systems is
to design distributed protocols based only on the local relative information to guarantee the states
of all agents to reach an agreement, known as the consensus problem. According to convergence rate, existing
consensus algorithms can be roughly categorized into two classes, namely, asymptotic consensus \cite{Song10,Su11,Hu12,Meng:13,Fu:14} and finite-time consensus \cite{Cortes06,Hui09,Wang08,Li11}. For the consensus control problem, a key
task is to design appropriate distributed controllers which are usually
called consensus protocols. Due to the practical engineering requirements of networked agents, designing finite-time consensus protocols has been a hot research topic in the area of consensus problem.

Previous works of study the
finite-time consensus problem was firstly shown in \cite{Cortes06} for first-order multi-agent systems, where the signed gradient flows of a differential function and discontinuous algorithms were used. Since then, a variety of finite-time consensus algorithms have been proposed to solve the finite-time consensus problem under different scenarios; see  \cite{Hui09,Wang08,Chen11} and references therein. Then, over directed graphs having a spanning tree, the finite-time consensus protocol designing problems was studied in \cite{Sayyaadi11,Mauro13,Caoauto14}.
Further, a class of finite-time consensus protocols for second-order multi-agent systems are given in \cite{Zhang13,Zhao13,Li11,Xu:13}. Then, the finite-time consensus problems for multiple non-identical second-order nonlinear systems in \cite{Zhao16} with the settling time estimation, where the settling time functions is depend on initial states of the agents. It prohibits
their practical applications if the knowledge of initial conditions is
unavailable in advance. Recently, the authors in \cite{Zuo12,Zuo15} present a
novel class of nonlinear consensus protocols for integrator-type
multi-agent networks, called fixed-time consensus which assumes
uniform boundedness of a settling time regardless of the initial
conditions. Also, for multiple linear systems, fixed-time formation problems were studied in \cite{Liu15}. However, most of the above-mentioned works in \cite{Zuo12,Zuo15,Liu15} are derived for multi-agent systems under undirected topologies. It is significant and challenging to solve the fixed-time consensus problem for linear  multi-agent systems over directed topologies.

Motivated by the above observations, by using motion
planning approaches, this technical note investigates the fixed-time
consensus problem of general linear multi-agent systems over directed graphs. Main contributions of this technical note can be summed up in the following aspects.
Firstly, by using motion planning approaches, a novel framework is introduced
to solve fixed-time consensus problems. In this framework, for general linear multi-agent systems considered in this technical note, a novel class of distributed protocols are designed to solve fixed-time consensus problems. In particular, the  settling time can be off-line pre-specified according to task requirements. Secondly,  the communication topologies among the networked agents in this technical note is directed and contains a directed spanning tree. To the best of authors' knowledge, it is the first time to solve fixed-time consensus problems for general linear multi-agent systems over directed graphs. Finally, the protocols designed in this technical note are based only on sampling measurements of the relative state information among its neighbors, which greatly reduces cost of the network communication.

\emph{Notations}: Let $R^n$ and $R^{n\times n}$ be the sets of real numbers and real matrices, respectively. $I_n$ represents the identity matrix of dimension $n$. Denote by $\mathbf{1}$ a column vector with
all entries equal to one. The matrix inequality $A> (\geq) B$ means that
$A-B$ is positive (semi-) definite. Denote by $A\otimes B$ the Kronecker product
of matrices $A$ and $B$. For a vector $x\in R^n$, let $\|x\|$
denote 2-norm of $x$. For a set $V$, $|V|$ represents the number of elements in $V$.

\section{Notation and Preliminaries}

Let $\mathbb{R}^{n\times n}$ be the set of $n\times n$ real matrices. The superscript $T$ means transpose for real matrices. $I_n$ represents the identity matrix of dimension $n$, and $I$
denotes the identity matrix of an appropriate dimension. Let $\mathbf{1}_{n}$ denote the vector with all entries equal to one. For $\xi\in \mathbb{C}$, $\mathrm{Re}(\xi)$ denote by its real part. $\mathrm{diag}(A_1,\cdots,A_N)$ represents a block-diagonal
matrix with matrices $A_i,\;i=1,\cdots,n$, on its diagonal. The
Kronecker product \cite{matrix} of matrices $A\in \mathbb{R}^{m\times n}$ and $A\in \mathbb{B}^{p\times q}$ is defined as
\begin{eqnarray*}
A\otimes B=\left[
             \begin{array}{ccc}
               a_{11}B & \ldots & a_{1n}B \\
               \vdots & \ddots & \vdots \\
               a_{m1}B & \ldots & a_{mn}B \\
             \end{array}
           \right]
\end{eqnarray*}
which satisfies the following properties:
\begin{eqnarray*}
(A\otimes B)(C\otimes D)=(AC\otimes BD),\\
(A\otimes B)^T=A^T\otimes B^T,\\
A\otimes B+A\otimes C=A\otimes (B+C).
\end{eqnarray*}

For systems with $n$ agents, a directed graph $\mathcal{G}=(\mathcal{V},\mathcal{E})$ is developed to model the interaction among these agents, where $\mathcal{V}=\{1,2,\cdots,N\}$ is the vertex set and $\mathcal{E}\subset \{(v_i,v_j):v_i,v_j\in \mathcal{V}\}$ is the edge set, where an edge is an ordered pair of vertices in $\mathcal{V}$ which means that agent $j$ can receive information from agent $i$. If there is a directed edge from $i$ to $j$ , $i$ is defined as the parent node and $j$ is defined as the child node. The neighbors of node $i$ are denoted by $\mathcal{N}_i = \{j \in \mathcal{V} | (v_i,v_j)\in \mathcal{E}\}$, and $|\mathcal{N}_i|$ is the neighbors number of agent $i$. A directed tree is a directed graph, where every node, except the root, has exactly one parent. A spanning tree of a directed
graph is a directed tree formed by graph edges that connect all the
nodes of the graph. We say that a graph has (or contains)
a spanning tree if a subset of the edges forms a spanning tree.

The adjacency matrix $A$ associated with $\mathcal{G}$ is defined such that $a_{ij}=1$ if node $i$ is adjacent to node $j$, and $a_{ij}=0$ otherwise. The Laplacian matrix of the graph associated with adjacency matrix $A$ is given as $\mathcal{L} = (l_{ij})$, where $l_{ii}=\sum\limits_{j =1}^{N} a_{ij}$ and $l_{ij} =-a_{ij}$, $i\neq j$.

\section{Distributed fixed-time consensus control for linear multi-agent systems over directed graphs}

In this section, the distributed fixed-time consensus problem for linear multi-agent systems over directed graphs is studied. Consider a multi-agent systems with $N$ agents with general linear dynamics, which may be regarded as the linearized model of some nonlinear systems \cite{Li10,Li13}. The dynamics of each agents in networks is described by
\begin{eqnarray}\label{1}
\dot x_i(t)=Ax_i(t)+Bu_i(t),\;\;\;i=1,2,\cdots,N,
\end{eqnarray}
where $x_i(t)=[x_{i1}(t),x_{i2}(t),\cdots,x_{in}(t)]^T\in \mathbb{R}^n$ is the state of agent $i$ , $u_i\in \mathbb{R}^m$ is its control input. $A$, $B$, are constant matrices with compatible dimensions, and $(A,B)$ is controllable.

\begin{assumption}\label{ass1}
Suppose that the graph $\mathcal{G}$ of the communication topology is directed and has a spanning tree.
\end{assumption}

%%%\begin{assumption}\label{ass2}
%%%$A$ has no eigenvalues with negative real parts.
%%%\end{assumption}

\begin{lemma} \label{lemma1}
\cite{Godsil}
Assume that directed graph $\mathcal{G}$ has a spanning tree. Then, zero is a simple eigenvalue of $\mathcal{L}$ with
$\mathbf{1}$ as an eigenvector and all of the nonzero eigenvalues are in the open right
half plane.
\end{lemma}
\begin{lemma}\label{lemma2}
\cite{Ren05}
Let $M=[m_{ij}]\in M_N(\mathbb{R})$ be a stochastic matrix, where $M_N(\mathbb{R})$ represents the set of all $N\times N$ real matrices. If $M$ has an eigenvalue $\lambda = 1 $ with algebraic multiplicity equal to one, and all the other eigenvalues satisfy $|\lambda| < 1$, then $M$ is SIA, that is, $\lim_{k\to \infty} M^k= \mathbf{1}\xi^T$ , where $\xi=(\xi_1,\xi_2,\cdots,\xi_N)^T\in R^N$ satisfies $M^T\xi=\xi$ and $\mathbf{1}^T\xi=1$.
Furthermore, each element of $\xi$ is nonnegative.
\end{lemma}
%\begin{definition}
%The asymptotic consensus problem for systems (\ref{1}) is solved if and only if
%\begin{eqnarray*}
%\lim_{t\to\infty}\|x_i(t)-x_j(t)\|=0,\;\forall i,j\in\mathcal{V},
%\end{eqnarray*}
%for any initial conditions.
%\end{definition}
\begin{definition}
The fixed-time consensus problem for multi-agent systems (\ref{1}) is said to be solved if and only if for a off-line pre-specified finite settling time $T_s>0$, states of multi-agent systems (\ref{1}) satisfy
\begin{eqnarray*}
\lim_{t\to T_s}\|x_i(t)-x_j(t)\|=0,\;\forall i,j\in\mathcal{V},
\end{eqnarray*}
and $x_i(t)=x_j(t)$ when $t>T_s$ for any initial conditions.
\end{definition}
In this technical note, it is assumed that only relative measurements information can be used to develop the distributed consensus protocols. Moreover, the $i$th agent can only obtain the consensus error from its neighborhood. The objective of this technical note is to design a distributed control law $u_i$ based on the above-mentioned relative information such that the states of all the agents in networks reach fixed-time consensus over directed graphs. In order to achieve the control objective in this technical note, the following sampled-type protocol is proposed:
\begin{eqnarray}\label{2}
u_i(t)=-\frac{1}{|\mathcal{N}_i|+1}B^TP\sum\limits_{j\in\mathcal{N}_i}\big(x_i(t_k)-x_j(t_k)\big),
\end{eqnarray}
where
\begin{eqnarray*}
   &&P=e^{-A^T(t-t_k)}\Phi^{-1} e^{A(t_{k+1}-t_k)},\\
&&\Phi
=\left(
                               \begin{array}{cc}
                                 I_n & 0_{n} \\
                               \end{array}
                             \right)
e^{M(t_{k+1}-t_k)}\left(
                    \begin{array}{c}
                      0_{n} \\
                      I_n \\
                    \end{array}
                  \right),\\
&&M=\left(
                    \begin{array}{ccc}
                        A & BB^T\\
                        0_{n\times n} & -A^T \\
                      \end{array}
                  \right),\\
                  &&t_k\leq t<t_{k+1}  ,\;\;i=1,2,\cdots,N,
\end{eqnarray*}
with the sampling time sequence $\{t_k|t_k=t_0+T_k, T_k=\frac{6}{(\pi k)^2}T_s\}$, where $T_s>0$ is a off-line pre-specified settling time according to task requirements.

The above fixed-time consensus protocol is designed by considering the following Hamiltonian function:
\begin{eqnarray} \label{4}
H_{i,k} = -\frac{1}{2}\sum\limits_{i =1}^{N}u_i^T(t)u_i(t) + \sum\limits_{i =1}^{N}p^T_i(t) (Ax_i(t)+Bu_i(t)),
\end{eqnarray}
where $p_i(t)\in \mathbb{R}^{n}$ represents the costate. Then,
(\ref{4}) is the Hamiltonian function of the cost function proposed as follows:
\begin{eqnarray} \label{5}
J_{i,k} = \frac{1}{2}\int_{t_k }^{t_{k+1}} \sum\limits_{i =1}^{N}{u_i^T(t) R_iu_i(t)dt},\qquad\qquad
\end{eqnarray}
where $t_k$ and $t_{k+1}$ can be  seen as the initial and terminal times, respectively.
Then, according to
Pontryagin's principle  \cite{optimal} that the necessary
condition of optimality are written as
\begin{subequations}\label{6}
\begin{numcases}{}
\dot x_i(t) = \frac{\partial H_{i,k}}{\partial p_i(t)}=Ax_i(t)+Bu_i(t),\label{6a} \\
\dot p_i(t) =-\frac{\partial H_{i,k}}{\partial x_i(t)}=-A^Tp_i(t).\label{6b}
\end{numcases}
\end{subequations}
Besides, according to the extremal condition
\begin{eqnarray*}
\frac{{\partial H_{i,k}}}{{\partial u_i(t)}} = -u_i(t) + B^Tp_i(t)  = 0,
\end{eqnarray*}
one has
\begin{eqnarray}\label{7}
u_i(t)=  B^T p_i(t).
\end{eqnarray}
Therefore, the determination of the optimal control (\ref{7}) is boiled down to computing $p_i(t)$.

By substituting (\ref{7}) into (\ref{6a}) and (\ref{6b}), one gets
\begin{eqnarray}\label{8}
\left(
  \begin{array}{c}
    \dot x_i (t) \\
    \dot p_i (t) \\
  \end{array}
\right)=\left(
          \begin{array}{cc}
            A & BB^T \\
            0_{n\times n} & -A^T \\
          \end{array}
        \right)\left(
                 \begin{array}{c}
                   x_i (t) \\
                   p_i (t) \\
                 \end{array}
               \right).
\end{eqnarray}
Integrating the above equation form $t_k$ to $t_{k+1}$, it follows that
\begin{eqnarray}\label{9}
\left(
                 \begin{array}{c}
                   x_i (t_{k+1}) \\
                   p_i (t_{k+1}) \\
                 \end{array}
               \right)&=&e^{M(t_{k+1}-t_k)}\left(
                 \begin{array}{c}
                   x_i (t_{k}) \\
                   p_i (t_{k}) \\
                 \end{array}
               \right).
\end{eqnarray}
Further, by designing the terminal condition of (\ref{4}) over $[t_k, t_{k+1}]$ as follows,
\begin{eqnarray}\label{10}
x_i(t_{k+1})&=&\frac{1}{|\mathcal{N}_i|+1}e^{A(t_{k+1}-t_k)}[\sum\limits_{j\in\mathcal{N}_i}x_j(t_k)+x_i(t_k)],\nonumber\\
&&\;\;\;i=1,2,\cdots,N,
\end{eqnarray}
from (\ref{9}) and (\ref{10}), one has
\begin{eqnarray*}
&&e^{A(t_{k+1}-t_k)}\frac{1}{|\mathcal{N}_i|+1}[\sum\limits_{j\in\mathcal{N}_i}x_j(t_k)+x_i(t_k)]\nonumber\\
&=&e^{A(t_{k+1}-t_k)}x_i(t_k)+ \Phi p_i(t_k).
\end{eqnarray*}
If $\Phi$ is invertible, then, it is followed that
\begin{eqnarray*}
p_i(t_k)=\Phi^{-1}e^{A(t_{k+1}-t_k)}\frac{1}{|\mathcal{N}_i|+1}\sum\limits_{j\in\mathcal{N}_i}[x_j(t_k)-x_i(t_k)]
\end{eqnarray*}
Therefore, for the time sequence ${t_k}$, one has the distributed protocol (\ref{2}).

To summarize so far, with the fixed-time control protocol (\ref{2}), the states of linear multi-agent systems (\ref{1}) are derived from $x_i(t_k)$ to $\frac{\sum\limits_{j\in\mathcal{N}_i}x_j(t_k)+x_i(t_k)}{|\mathcal{N}_i|+1},\;i=1,2,\cdots,N$. That is to say, with the distributed controller (\ref{2}), the states of each agent in systems (\ref{1}) will be derived to the average state of all its neighbors. From an intuitional point of view, after enough times such motion planning steps, the states of all agents in systems (\ref{1}) will achieve consensus.

Note that, the above proposed protocol exists only if $\Phi$ is invertible. Thus, before moving on, the following lemma is given.

\begin{lemma}\label{lemma3}
$\Phi$ is invertible if and only if $(A,B)$ is controllable.
\end{lemma}
\textbf{Proof}: It follows from (\ref{2}) that
\begin{eqnarray*}
\Phi
&\!\!\!\!\!{=}\!\!\!\!\!&I_N{\otimes}\big((t_{k+1}{-}t_k)BB^T{+}\frac{(t_{k+1}{-}t_k)^2}{2!}(ABB^T{-}BB^TA^T)\nonumber\\
&&{+}\frac{(t_{k+1}-t_k)^3}{3!}(A^2BB^T{-}ABB^TA^T{+}BB^T{A^T}^2)\nonumber\\
&&{+}\frac{(t_{k+1}-t_k)^4}{4!}(A^3BB^T{-}A^2BB^TA^T{+}ABB^T{A^T}^2\nonumber\\
&&{-}BB^T{A^T}^3){+}\cdots\cdots\big).\qquad\qquad
\end{eqnarray*}
First, the proof of the sufficiency is given as follows.
Let
\begin{eqnarray*}
\Pi&=&tBB^T+\frac{t^2}{2!}(ABB^T{-}BB^TA^T)+\frac{t^3}{3!}(A^2BB^T
\\
&&-ABB^TA^T+BB^T{A^T}^2)+\frac{t^4}{4!}(A^3BB^T\\
&&-A^2BB^TA^T
+ABB^T{A^T}^2-BB^T{A^T}^3)+\cdots\cdots.
\end{eqnarray*}
By reductio, assume that $\Pi$ is singular. Thus, there is at least one nonzero
vector $\alpha\in \mathbb{R}^n$, which makes that
\begin{eqnarray*}
\alpha^T\Pi=0.
\end{eqnarray*}
Taking the derivatives of the above equation to $(n-1)$ order with respect to the time $t$ and setting $t=0$, we get
\begin{eqnarray*}
&&\alpha^TBB^T{=}0,\;\\
&&\alpha^T(ABB^T{-}BB^TA^T){=}0,\;\\
&&\alpha^T(A^2BB^T-ABB^TA^T+BB^T{A^T}^2){=}0,\;\\
&&\alpha^T(A^3BB^T{-}A^2BB^TA^T{+}ABB^T{A^T}^2{-}BB^T{A^T}^3){=}0,\\
&&\qquad\qquad\qquad\qquad\qquad\cdots\cdots\\
&&\alpha^T(A^{n-1}BB^T{-}A^{n-2}BB^TA^T{+}A^{n-3}BB^T{A^T}^2{-}\\
&&\;\;\;\cdots{+}(-1)^{n-1}BB^T{A^T}^{n-1}){=}0.
\end{eqnarray*}
Simplifying the above equations, we have
\begin{eqnarray*}
&&\alpha^TBB^T=0,\;\alpha^TABB^T=0,\;\alpha^TA^2BB^T=0,\;\\
&&\alpha^TA^3BB^T=0,\;\cdots\cdots,\alpha^TA^{n-1}BB^T=0.
\end{eqnarray*}
Let
\begin{eqnarray*}
Q{=}\left[
    \begin{array}{cccccc}
      BB^T & ABB^T & A^2BB^T &  \cdots & A^{n{-}1}BB^T\\
    \end{array}
  \right].
\end{eqnarray*}
Then, we have $\alpha^TQ=0$.
It follows from $\alpha \neq 0$ that the matrix $Q$ is linearly dependent. Note that $(A,BB^T)$ is controllable, if and only if $(A,B)$ is controllable. This contradicts with the condition that $(A,B)$ is controllable.
Thus, it is proved that $\Pi$ is nonsingular. Further, $\Phi$ is invertible.

Similarly, the proof of the necessity is given. By reductio, it is assumed that $(A,B)$ is uncontrollable. Thus, there is at least one nonzero vector $\beta\in \mathbb{R}^n$, which makes that
\begin{eqnarray*}
\beta^TQ=0.
\end{eqnarray*}
It follows that
\begin{eqnarray*}
&&\beta^TBB^T=0,\;\beta^TABB^T=0,\;\beta^TA^2BB^T=0,\;\\
&&\beta^TA^3BB^T=0,\;\cdots\cdots,\beta^TA^{n-1}BB^T=0.
\end{eqnarray*}
Then, it is obtained that
\begin{eqnarray*}
\beta^T\Pi=0.
\end{eqnarray*}
This contradicts with the condition that $\Phi$ is invertible.
Thus, it is proved that $(A,B)$ is controllable. The proof is completed.
{\remark{From the Lemma \ref{lemma3}, one has that the the proposed protocol exists if $(A,B)$ is controllable, which is a fundamental requirement for control of linear systems.}}

{\remark{Existing works \cite{Li10} usually design consensus protocols using the smallest real part of the nonzero eigenvalues of the Laplacian matrix associated
with the communication graph. Some researchers have used the adaptive control approaches \cite{Li13} to overcome the requirements of  nonzero eigenvalues of the Laplacian matrix. In this technical note, the above proposed protocol (\ref{2}) does not require the knowledge of the Laplacian matrix associated
with the communication graph.}}

%\begin{remark}
%It is worth mentioning that the above sampled-data consensus protocol (\ref{2}) is
%designed based on motion planing approaches. Specifically, consider the cost function $J_k = \frac{1}{2}\int_{t_k }^{t_{k+1}} \sum\limits_{i =1}^{N}{u_i^T(t) R_iu_i(t)dt}$ and the associated Hamiltonian function
%$
%H_k(t) = -\frac{1}{2}\sum\limits_{i =1}^{N}u_i^T(t)u_i(t) + \sum\limits_{i =1}^{N}p^T_i(t) (Ax_i(t)+Bu_i(t)), t_k\leq t<t_{k+1}, k=0,1,\cdots,$
%with terminal conditions $$x_i(t_{k+1})=e^{A(t_{k+1}-t_k)}\frac{1}{|\mathcal{N}_i|+1}[\sum\limits_{j\in\mathcal{N}_i}x_j(t_k)+x_i(t_k)],$$ $$\;\;\;i=1,2,\cdots,N,$$
%where $p_i(t)\in \mathbb{R}^{n}$ represents the costate. Solve the above optimal planing problem in light of Pontryagin's principle \cite{optimal}. One obtains the proposed protocol (\ref{2}) in this section.
%From the terminal conditions above, one has that protocol (\ref{2}) is designed in order to derive the state of every agent in systems (\ref{1}) to the average states of all its neighbors at sampling instants. From an intuitional point of view, after several times such motion planning, the states of networked agents in systems (\ref{1}) will achieve consensus.
%To verify this viewpoint, the following theorem is given.
%\end{remark}

Then, the following theorem provides the main result in this technical note.

\begin{theorem}
Suppose that Assumption \ref{ass1} holds. For an off-line pre-specified  settling time
$T_s$, the distributed sampling protocol (\ref{2}) can solve the fixed-time
consensus problem of linear multi-agent system (\ref{1}) if $(A,B)$ is controllable.
%if the sampling time $\{T_k=t_{k+1}-t_k,\;k=1,2,\cdots\}$ is upper bounded, and the following condition is satisfied
%\begin{eqnarray}\label{3}
%(1-\frac{\lambda_{2}}{|\mathcal{N}|+1})e^{Re(\nu_{max})(t_{k+1}-t_k)}<1,
%\end{eqnarray}
%where $|\mathcal{N}|=\max_{i=1,\cdots,N}|\mathcal{N}_i|$, $\lambda_{2}$ is the smallest nonzero eigenvalue of $\mathcal{L}$, and $\nu_{max}$ is the eigenvalue of $A$ with the largest real part.
\end{theorem}
\textbf{Proof}:
First, it is to prove that at the sampling time series $\{t_k\}$, the states of systems (\ref{1}) with (\ref{2}) will achieve consensus.
Substituting (\ref{2}) into (\ref{1}), one gets
\begin{eqnarray}\label{closedloopsystem}
\dot x_i(t)&=&Ax_i(t)-\frac{1}{|\mathcal{N}_i|+1}BB^Te^{-A^T(t-t_k)}\nonumber\\
&&\cdot \Phi^{-1} e^{A(t_{k+1}-t_k)}\sum\limits_{j\in\mathcal{N}_i}\big(x_i(t_k)-x_j(t_k)\big).
\end{eqnarray}
By integrating (\ref{closedloopsystem}) from $t_k$ to $t_{k+1}$, $k=0,1, \cdots$, one gets
\begin{eqnarray*}
&&x_i(t_{k+1})-e^{A(t_{k+1}-t_k)}x_i(t_{k})\\
&&=e^{A(t_{k+1}-t_k)}\frac{1}{|\mathcal{N}_i|+1}\sum\limits_{j\in\mathcal{N}_i}\big(x_j(t_k)-x_i(t_k)\big).
\end{eqnarray*}
Let $X(t_{k})=(x^T_1(t_{k}),x^T_2(t_{k}),\cdots,x^T_N(t_{k}))^T$.
It follows that
\begin{eqnarray*}
&&X(t_{k+1})-I_N\otimes e^{A(t_{k+1}-t_k)}X(t_{k})\\
&&=-(\mathcal{N}\otimes e^{A(t_{k+1}-t_k)}) (\mathcal{L}\otimes I_n)\cdot X(t_{k}),
\end{eqnarray*}
where $\mathcal{N}=\mathrm{diag}(\frac{1}{|\mathcal{N}_1|+1}, \frac{1}{|\mathcal{N}_2|+1},\cdots, \frac{1}{|\mathcal{N}_N|+1})$.
Then,
\begin{eqnarray*}
X(t_{k+1})&=&(I_N-\mathcal{N}\mathcal{L})\otimes e^{A(t_{k+1}-t_k)}X(t_{k})\\
&=&(I_N-\mathcal{N}\mathcal{L})^{k+1}\otimes e^{A(t_{k+1}-t_0)}X(t_{0}).
\end{eqnarray*}
Under Assumption \ref{ass1}, the directed graph $\mathcal{G}$ has a spanning tree. Thus, $I_N-\mathcal{N}\mathcal{L}$ is a stochastic matrix. According to Lemma \ref{lemma1}, one gets that $I_N-\mathcal{N}\mathcal{L}$ has an eigenvalue $\lambda_1 = 1 $ with algebraic multiplicity equal to one, and all the other eigenvalues satisfy $|\lambda_i| < 1,\;i=2,\cdots,N$. Thus, it followed from Lemma \ref{lemma2} that for matrix $I_N-\mathcal{N}\mathcal{L}$, there exist a column vector $\xi$ such that
\begin{eqnarray}\label{a}
\lim_{k\to\infty}(I_N-\mathcal{N}\mathcal{L})^{k}=\mathbf{1}\mathbf{\xi}^T.
\end{eqnarray}
Besides, since $t_{k}\to T_s$ as $k\to \infty$, one has $t_{k-1}{-}t_{0}$ is bounded, which ensures each item of matrix $e^{A(t_{k+1}-t_0)}$ is bounded. It follows that
\begin{eqnarray}\label{b}
\lim_{k{\to} \infty}[(I_N{-}\mathcal{N}\mathcal{L})^{k}{-}\mathbf{1}\mathbf{\xi}^T]\otimes e^{A(t_{k+1}-t_0)} =0.
\end{eqnarray}
Let $x^\ast(t)
=\sum_{i=1}^N\xi_i e^{A(t-t_0)}x_i(t_{0})$ and $X^\ast(t)=\mathbf{1}\otimes x^\ast(t) $.
One has
\begin{eqnarray*}
X^\ast(t)
&=&[\mathbf{1}\mathbf{\xi}^T\otimes e^{A(t-t_0)}]X(t_{0}).
\end{eqnarray*}
Thus,
\begin{eqnarray*}
&&\lim_{k{\to} \infty}(X(t_{k+1})-X^\ast(t_{k+1}))\\
&=&\lim_{k{\to} \infty}[((I_N-\mathcal{N}\mathcal{L})^{k+1}-\mathbf{1}\mathbf{\xi}^T)\otimes e^{A(t_{k+1}-t_0)}]X(t_{0})\\
&=&0.
\end{eqnarray*}
Note that
\begin{eqnarray*}
X^\ast(t_k)
&=&\mathbf{1}\otimes\bigg[\sum_{i=1}^N\xi_i e^{A(t_k-t_0)}x_i(t_{0})\bigg].
\end{eqnarray*}
Thus, one has the discrete states $x_i(t_{k})$ will achieve consensus in exponential rate as $k\to \infty$, i.e., $\lim_{k\to\infty}\|x_i(t_{k})-x_j(t_k)\|= 0$.\\
Secondly, for the off-line pre-specified finite-time $T_s$, we will proof that the discrete states $x_i(t_k)$ can achieve fixed-time consensus as $t_k\to T_s$.
According to  $\{t_k|t_{k+1}=t_k+T_{k+1}, T_k=\frac{6}{(\pi k)^2}T_s, \;k=1,2,\cdots\}$, one has $\lim_{k\to\infty}t_{k}= T_s$. Thus,
\begin{eqnarray*}
&&\lim_{t_k\to T_s}\|x_i(t_{k})-x_j(t_k)\|\\
&=&\lim_{k\to \infty}\|x_i(t_{k})-x_j(t_k)\|\\
&=&0,i,j=1,2,\cdots,N
\end{eqnarray*}
Therefore, for the pre-specified settling time $T_s$, the distract states $x_i(t_{k})$
will achieve fixed-time consensus in exponential rate as $t_k\to T_s$.\\
%
%\begin{eqnarray*}
%\Gamma_i(t_{k})=\mathcal{L}\otimes I_n\cdot X_i(t_{k}).
%\end{eqnarray*}
%One has
%\begin{eqnarray}\label{11}
%\Gamma_i(t_{k+1})=(I_N-\mathcal{L}\mathcal{N})\otimes e^{A(t_{k+1}-t_k)}\cdot\Gamma_i(t_{k}).
%\end{eqnarray}
%Note that the above system is asymptotic stable if all eigenvalues of $(I_N-\mathcal{L}N)\otimes e^{A(t_{k+1}-t_k)}$ lie within the open
%unit circle.
%One has the following condition
%\begin{eqnarray*}
%\max_{i=1,\cdots,N,\;j=1,\cdots,n}{(1-\frac{\lambda_i}{|\mathcal{N}|+1})e^{Re(\nu_j)(t_{k+1}-t_k)}}<1,
%\end{eqnarray*}
%where $\lambda_i$ and $\nu_j$ are the eigenvalues of $\mathcal{L}$ and $A$, respectively. Therefore, system (\ref{11}) is asymptotic stable if the condition (\ref{3}) is satisfied, i.e.
%\begin{eqnarray*}
%(1-\frac{\lambda_{2}}{|\mathcal{N}|+1})e^{Re(\nu_{max})(t_{k+1}-t_k)}<1.
%\end{eqnarray*}
%Considering the property of $\mathcal{L}$, one has $\lim_{k\to\infty}x_i(t_{k})-x_j(t_k)\rightarrow 0, i,j=1,2,\cdots,N$.
Finally, we will proof that the the continuous states $x_i(t)$ can achieve fixed-time consensus as $t\to T_s$.
By integrating (\ref{closedloopsystem}) from $t_k$ to $t$, it is obtained that
\begin{eqnarray*}
&&x_i(t)-e^{A(t-t_k)}x_i(t_k)\\
&{=}&{-}\frac{1}{|\mathcal{N}_i|+1}\int_{t_k }^{t}e^{A(t{-}\tau)}BB^Te^{{-}A^T(\tau-t_k)}d\tau \\
&&\cdot \Phi^{-1} e^{A(t_{k+1}-t_k)}\sum\limits_{s\in\mathcal{N}_i}\big(x_i(t_k)-x_s(t_k)\big),\\
&&\;t_k\leq t<t_{k+1}.
\end{eqnarray*}
Let
\begin{eqnarray*}
\delta_i(t)
&=
&-\frac{1}{|\mathcal{N}_i|+1}\int_{t_k }^{t}e^{A(t{-}\tau)}BB^Te^{{-}A^T(\tau-t_k)}d\tau\\
&&\cdot\Phi^{-1} e^{A(t_{k+1}-t_k)}\sum\limits_{s\in\mathcal{N}_i}\big(x_i(t_k)-x_s(t_k)\big).
\end{eqnarray*}
Thus,
\begin{eqnarray*}
x_i(t)-x_j(t)=e^{A(t-t_k)}(x_i(t_k)-x_j(t_k))+\delta_i(t)-\delta_j(t).
\end{eqnarray*}
Further,
\begin{eqnarray*}
\|x_i(t)-x_j(t)\|
&\leq&\parallel e^{A(t_{k+1}-t_k)}\parallel\cdot\parallel x_i(t_{k})-x_j(t_{k})\parallel\\
&&+\parallel\delta_i(t_k)\parallel+\parallel\delta_j(t_k)\parallel.
\end{eqnarray*}
Besides, note that
\begin{eqnarray*}
\parallel\delta_i(t)\parallel&\leq& \frac{\parallel e^{A(t_{k+1}-t_k)}\parallel}{|\mathcal{N}_i|+1} \parallel\int_{t_k }^{t_{k+1}}BB^Te^{-A^T(\tau-t_k)}d\tau \parallel\\
&&\cdot\parallel\Phi^{-1}\parallel\parallel e^{A(t_{k+1}-t_k)}\parallel\\
&&\cdot\sum\limits_{s\in\mathcal{N}_i}\parallel x_i(t_k)-x_s(t_k)\parallel.
\end{eqnarray*}
Since the length of the time interval $t_k-t_{k-1}=T_k=\frac{6}{(\pi k)^2}T_s,\;k=1,2,\cdots\}$ is upper bounded. One has $\parallel e^{A(t_{k+1}-t_k)}\parallel$ and $\parallel\int_{t_k }^{t_{k+1}}BB^Te^{-A^T(\tau-t_k)}d\tau \parallel$ are bounded. Furthermore, since $(A, B)$ is controllable, it follows from Lemma \ref{lemma3} that $\Phi$ is invertible and $\Phi^{-1}=\frac{\Phi^\ast}{|\Phi|}$, where ${|\Phi|}\neq 0$. Thus, by assuming that $|\Phi|=\sum_{s=0}^\infty f_s(t_{k+1}-t_k)^s$, where $f_s, s=0,1,\cdots$ are coefficients, there exists at least a finite constant $m$ such that coefficient $f_m\neq 0$ and $f_0=f_1=\cdots=f_{m-1}= 0$. Thus, $|\Phi|$ can be rewritten as $|\Phi|= f_m(t_{k+1}-t_k)^m+\circ(t_{k+1}-t_k)$ when $t_{k+1}-t_k$ is very small, where $\circ(t_{k+1}-t_k)$ presents the higher-order infinitesimal of $t_{k+1}-t_k$. Since $\lim_{k\to \infty}(t_k-t_{k-1})=\lim_{k\to \infty}T_k=\lim_{k\to \infty}\frac{6}{(\pi k)^2}T_s=0$ in polynomial rate, one has
$\lim_{k\to \infty}\frac{\parallel x_i(t_{k})-x_j(t_k)\parallel}{|\Phi|} = 0$. According that $\lim_{k\to \infty}t_k=T_s$, therefore,
$\lim_{t_k\to T_s}\parallel\Phi^{-1}\parallel \parallel x_i(t_{k})-x_j(t_k)\parallel= 0$. It is obtained that $\lim_{t\to T_s}\parallel\delta_i(t)\parallel= 0 $. Similarly, one gets $\lim_{t\to T_s}\parallel\delta_j(t)\parallel= 0 $. Therefore, $\lim_{t\to T_s}\|x_i(t)-x_j(t)\|= 0$.

To sum up, under the protocol (\ref{2}), linear multi-agent systems (\ref{1}) can achieve consensus in the pre-specified fixed settling time $T_s$. The proof is completed.
\begin{remark}
Under undirected graphs, finite-time and fixed-time consensus problems have been investigated in some interesting papers \cite{Hui09,Wang08,Li11,Zhao13,Zuo15,Liu15,Zhao16,Xu:13}.
Under directed graphs, the finite-time consensus problem of first-order multi-agent systems has been solved in \cite{Sayyaadi11,Mauro13,Caoauto14}. However, the algorithms in \cite{Sayyaadi11,Mauro13,Caoauto14} are difficult to develop for solving the finite-time consensus problem of high-order multi-agent systems under directed graphs. In this technical note,
by using motion planning approaches, the designed protocol  \ref{2} successfully solves the fixed-time consensus problems for general linear multi-agent systems over directed graphs.
\end{remark}

\begin{remark}
Compared with the existing works \cite{Li11,Hui09,Zhang13,Zhao16} on finite-time consensus problems, in this article, the settling time can be off-line pre-specified according to task requirements, which not only realizes the consensus in the state space but also controls the settling time in the time axis.
\end{remark}
\begin{remark}
It is worth of mentioning that the fixed-time protocols designed in this technical note are
based only on sampling measurements of the relative state information among its neighbors, which greatly reduces cost of the network communication \cite{Yu11,Huang16,Wen13}.
\end{remark}

{\remark{On connected undirected graphs, fixed-time consensus algorithms have been studied in \cite{Zuo12,Zuo15}. As a special case of the result in Theorem 1 of this paper, on directed graphs containing a directed spanning tree, the fixed-time consensus problems for the multi-agent systems with dynamics such as single-integrator \cite{Zuo12}, double-integrator \cite{Zuo15} and harmonic oscillators \cite{Zhang13} can be solved in this technical note.}}

\section{Application to spacecraft formation flying.}
In this section, the application of the preceding control laws to spacecraft formation
flying in the low Earth orbit is addressed. Spacecraft formation flying needs precise coordination among multiple spacecraft whose dynamics are coupled through a common control law \cite{Scharf04}. Early pertinent works for spacecraft formation flying with variable dynamics, such as second integrators, linear systems, precise nonlinear models, have been launched by Ren \cite{Ren07}, Li \cite{Li10}, and so on, where control laws are developed for spacecraft to asymptotically convergent to the desired formation. In order to simplify the analysis, in this section, it is assumed that the reference orbit is a circular or near-circular orbit of radius $R_0$, i.e., $e_r=0$, where $e_r$ is the eccentricity of the reference orbit. The relative motion of the spacecraft with respect to the reference orbit can be described in the local vertical local horizontal (LVLH) frame. Let $\mathbf{r}=[x_,y,z]^T$ be the position vector of the spacecraft and $r=|\mathbf{r}|$. For the circular or near-circular reference orbit, the relative position dynamics of spacecraft with respect to the reference orbit can be written as given in \cite{Liu15}.
Further, assume that the relative orbit radius between the $i$th spacecraft and the reference orbit is very small compered to the radius of the reference orbit. The linearized equations of the relative dynamics of the $i$th spacecraft with respect to the reference orbit are given by Hill's equations
\begin{eqnarray}\label{28}
\ddot{x}-3n^2_rx-2n_r\dot y=u_{xi},\nonumber\\
\ddot{y}+2n_r\dot x=u_{yi},\nonumber\\
\ddot{z}+n^2_rz=u_{zi},
\end{eqnarray}
where $n_r$ is the natural frequency of the reference orbit, $u_{xi}$, $u_{yi}$ and $u_{zi}$ are the control inputs. Let $u_i=[u_{xi},u_{yi},u_{zi}]^T$. Thus, for the Hill' equations, one has the following form
\begin{eqnarray*}
\left[
    \begin{array}{ccc}
  \dot r_i \\
  \ddot r_i \\
\end{array}
  \right]=\left[
    \begin{array}{ccc}
                0 & I_3 \\
                A_1 & A_2 \\
              \end{array}
  \right]\left[
    \begin{array}{ccc}
  r_i \\
  \dot r_i \\
\end{array}
  \right]+\left[
    \begin{array}{ccc}
  0 \\
  I_3 \\
\end{array}
  \right]u_i,
\end{eqnarray*}
where
\begin{eqnarray*}
A_1{=}\left[
    \begin{array}{ccc}
      3{n_r}^2 & 0 & 0 \\
      0 & 0 & 0  \\
      0 & 0 & {-}{n_r}^2
    \end{array}
  \right],
A_2{=}\left[
              \begin{array}{ccc}
             0& 2{n_r}  & 0\\
      {-}2{n_r} & 0 & 0 \\
       0 & 0 & 0
              \end{array}
            \right].
\end{eqnarray*}

Spacecraft are said to achieve formation flying in fixed settling time if their velocity vectors converge to the same value and their positions maintain a prescribed separation, i.e., $r_i-h_i\rightarrow \dot r_j-h_j$, $\dot r_i\rightarrow \dot r_j$, $i,j=1,\cdots,N$, at the terminal time $T_s$, where $h_i-h_j\in \mathbb{R}^3$ denotes the desired constant separation between spacecraft $i$ and $j$, and $T_s$ can be given in advance. Based on algorithm (\ref{2}), the distributed fixed-time formation control law for spacecraft $i$ is proposed as
\begin{eqnarray}\label{30}
  u_i(t)\!\!\!\!\!&=&\!\!\!\!\!-A_1h_i-\frac{1}{|\mathcal{N}_i|+1}B^Te^{-A^T(t-t_k)}\Phi^{-1}e^{A(t_{k+1}-t_k)}\nonumber\\
   &&\cdot{\sum\limits_{j\in\mathcal{N}_i}\left[
    \begin{array}{ccc}
                r_i(t_k)-r_j(t_k)-h_i+h_j \\
                \dot r_i-\dot r_j\\
                \end{array}
            \right],}  \nonumber \\
   &&t_k\leq t<t_{k+1}  ,\;\;i=1,2,\cdots,N,
\end{eqnarray}
where $A=\left[
    \begin{array}{ccc}
                0 & I_3 \\
                A_1 & A_2 \\
             \end{array}
            \right]$, $B=\left[
    \begin{array}{ccc}
  0 \\
  I_3 \\
  \end{array}
            \right]$.
Note that  $(A,B)$ is controllable. Thus, according to Theorem 1, the following theorem is given.
\begin{theorem}
Assume that the directed topology graph $\mathcal{G}$ among the $N$ satellites satisfies Assumption \ref{ass1}. Then, for an off-line pre-specified settling time $T_s>0$, the fixed-time formation protocol (\ref{30}) with the sampling time sequence $\{t_k|t_k=t_0+T_k, T_k=\frac{6}{(\pi k)^2}T_s\}$
solves the fixed-time formation flying problem of multiple satellites systems described by (\ref{28}), i.e., $\lim_{t\to T_s}\|r_i(t)-r_j(t)+h_i-h_j\|=0,\; \lim_{t\to T_s}\|\dot r_i(t)-\dot r_j(t)\|=0,\;i,j=1,\cdots,N$.
\end{theorem}

{\bf Example}
Consider the formation flying of six spacecraft with respect to a circular reference orbit with the orbital radius $R_0=4.224\times 10^7 m$. Note that the gravitation constant of the earth $\mu=GM_e=3.986\times 10^{14} m^3/{s^2}$, where $G$ is the universal constant of gravity and $M_e$ is the mass of the Earth. Therefore, the natural frequency of the reference orbit $n_r=7.273\times 10^{-5} s^{-1}$. All the spacecraft have mass $m=410 kg$. The directed communication topology between the spacecrafts is given in Fig. 1. The desired formation is that the six satellites will maintain a regular hexagon with a separation of $1000 m$. Thus, it is given that $h_1=[0,1000,0]^T m, h_2=[-866,500,0]^T m, h_3=[-866,-500,0]^T m,
h_4=[0,-1000,0]^T m, h_5=[866,-500,0]^T m, h_6=[866,500,0]^T m$. To simplify things, let the initial time be $t_0=0 h$. The  off-line pre-specified  formation time is $200 h$. Select the initial states for spacecraft as following:
\begin{eqnarray*}
&&[r_1(t_0)]{=}[0,966000,10000]^T m,\;\;[\dot r_1(t_0)]{=}[10,0,0]^T m/s,\\
&&[r_2(t_0)]{=}[0,900000,20000]^T m,\;\;[\dot r_2(t_0)]{=}[15,0,0]^T m/s,\\
&&[r_3(t_0)]{=}[0,866000,30000]^T m,\;[\dot r_3(t_0)]{=}[20,0,0]^T m/s,\\
&&[r_4(t_0)]{=}[0,800000,40000]^T m,\;[\dot r_4(t_0)]{=}[25,0,0]^T m/s,\\
&&[r_5(t_0)]{=}[0,766000,50000]^T m,\;[\dot r_5(t_0)]{=}[30,0,0]^T m/s,\\
&&[r_6(t_0)]{=}[0,700000,60000]^T m,\;[\dot r_6(t_0)]{=}[35,0,0]^T m/s.
\end{eqnarray*}

\begin{figure}
\begin{center}
\includegraphics[height=5cm]{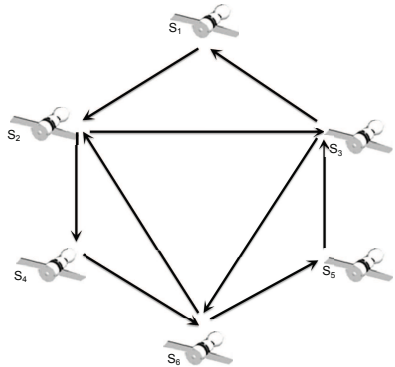}    % The printed column
\caption{Communication topology with 6 spacecrafts over a directed graph having a spanning tree.}  % width is 8.4 cm.
\label{fig1}                                 % Size the figures
\end{center}                                 % accordingly.
\end{figure}

\begin{figure}
\begin{center}
\includegraphics[height=6cm]{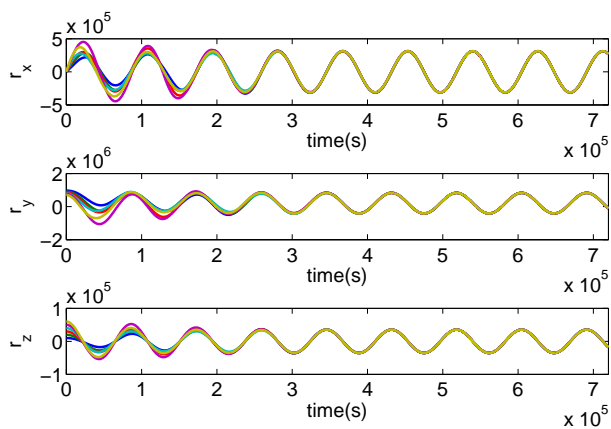}    % The printed column
\caption{The positions of the spacecrafts.}  % width is 8.4 cm.
\label{fig1}                                 % Size the figures
\end{center}                                 % accordingly.
\end{figure}

\begin{figure}
\begin{center}
\includegraphics[height=6cm]{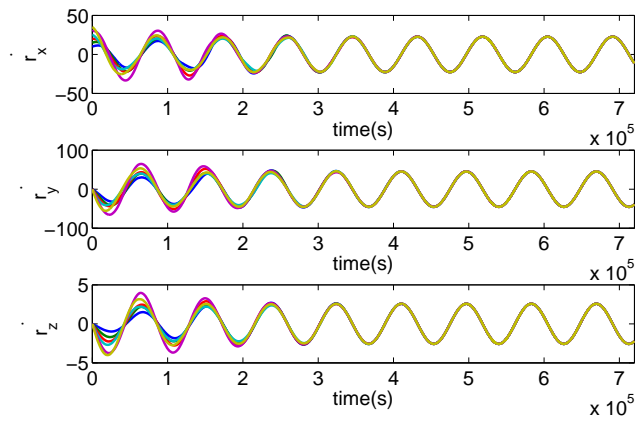}    % The printed column
\caption{The velocities of the spacecrafts.}  % width is 8.4 cm.
\label{fig1}                                 % Size the figures
\end{center}                                 % accordingly.
\end{figure}

\begin{figure}
\begin{center}
\includegraphics[height=6cm]{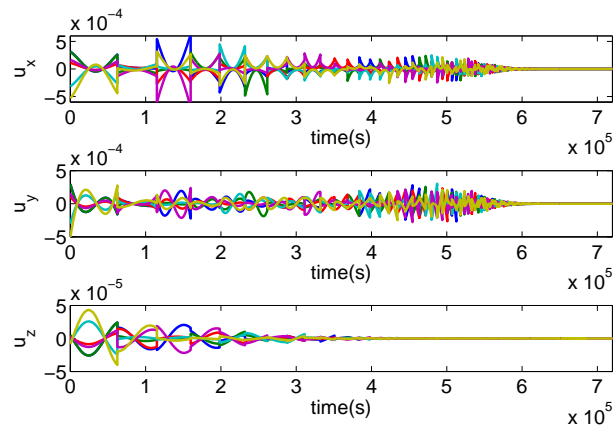}    % The printed column
\caption{The control signals of the spacecrafts.}  % width is 8.4 cm.
\label{fig1}                                 % Size the figures
\end{center}                                 % accordingly.
\end{figure}

\begin{figure}
\begin{center}
\includegraphics[height=5cm]{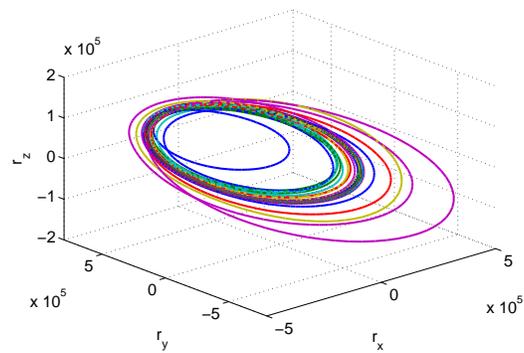}    % The printed column
\caption{The trajectories of the spacecrafts in three-dimensional space.}  % width is 8.4 cm.
\label{fig1}                                 % Size the figures
\end{center}                                 % accordingly.
\end{figure}

\begin{figure}
\begin{center}
\includegraphics[height=4cm]{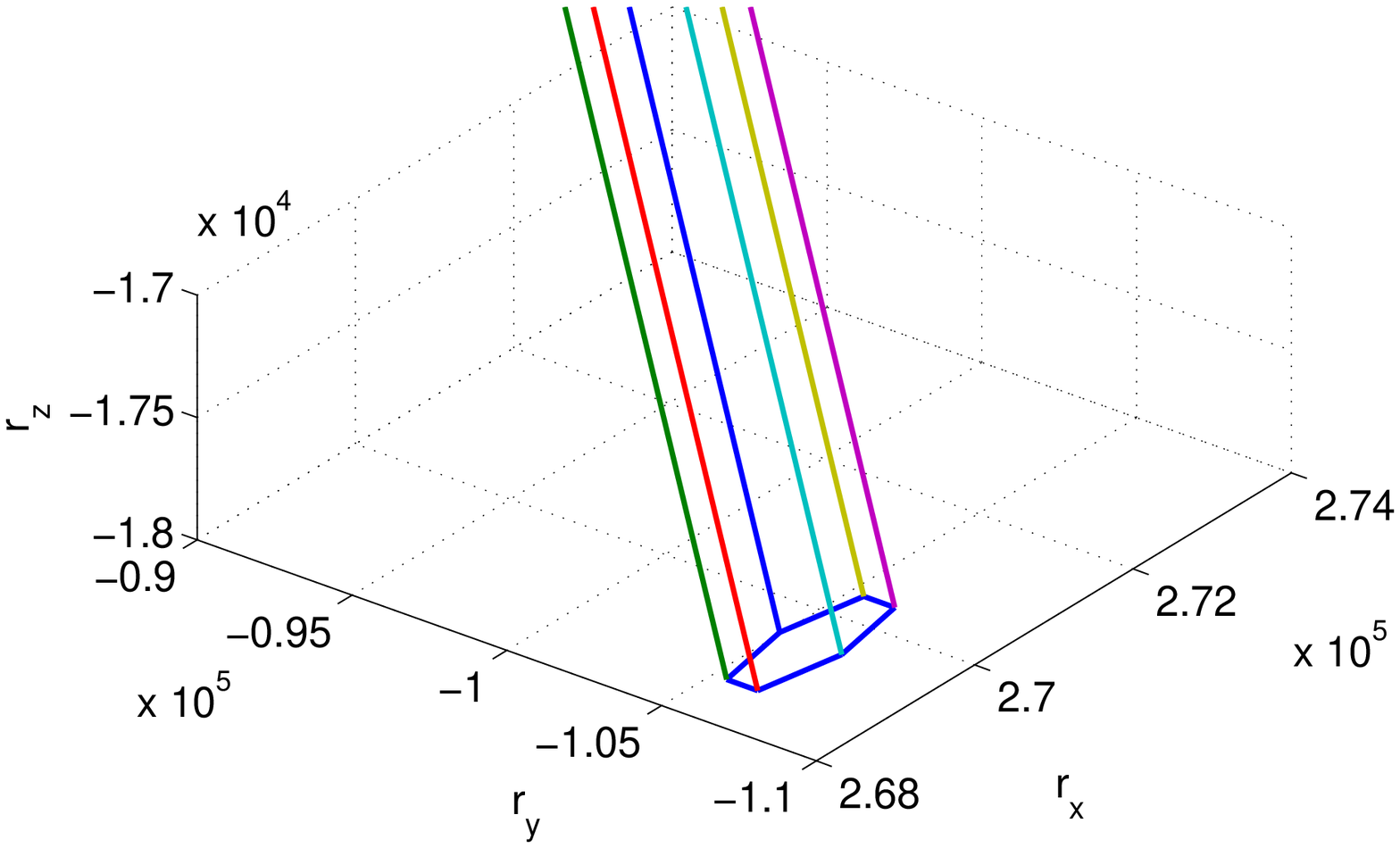}    % The printed column
\caption{The final formation
configurations of the spacecrafts in three-dimensional space.}  % width is 8.4 cm.
\label{fig1}                                 % Size the figures
\end{center}                                 % accordingly.
\end{figure}

Figs.2 and 3 depict the positions and the velocities of spacecrafts, respectively. The control forces are shown in Fig. 4. It can be seen that the desired formation is derived at the fixed settling time $T_s=200h$. The motion trajectories of these six spacecraft in three-dimensional space are illustrated in Fig.5. Fig. 6 shows the final formation
configurations. The six agents form a regular hexagon
about $1000m$ on each side.

\section{Conclusions}
This technical note has studied the distributed fixed-time consensus
protocol design problem for multi-agent systems with general continuous-time
linear dynamics over directed graphs. By using motion planning approaches, a class of distributed fixed-time consensus algorithms are developed, which rely only on the sampling information at some sampling instants. For linear multi-agent systems,  the proposed algorithms solve the fixed-time consensus problem for any directed graph containing a directed spanning tree. In particular, the fixed settling time can be off-line pre-specified according to task requirements.
Extensions to the fixed-time formation flying are further studied for multiple satellites described by Hill equations.
Future works will focus on solving distributed fixed-time
consensus problem for mobile agents modeled by nonlinear dynamics over directed switching graphs.

\bibliographystyle{IEEEtran}

\end{document}